\documentclass[12pt]{article}
\usepackage{a4,epsf}

\newcommand{\bea}{\begin{eqnarray}}
\newcommand{\eea}{\end{eqnarray}}
\newcommand{\simgt}{\hbox{ \raise3pt\hbox to 0pt{$>$}\raise-3pt\hbox{$\sim$} }}
\newcommand{\simlt}{\hbox{ \raise3pt\hbox to 0pt{$<$}\raise-3pt\hbox{$\sim$} }}

\begin{document}
\begin{titlepage}
\title{
\vspace{2cm}
Neutrino masses: hierarchy without hierarchy
}
\author{M.~Je\.zabek\\
\small
Henryk Niewodnicza\'nski Institute of Nuclear Physics,\\
\small
Kawiory 26a, PL-30055 Cracow, Poland
\\   \small and \\ 
\small
Institute of Physics, University of Silesia, \\
\small
     Uniwersytecka 4, PL-40007 Katowice, Poland\\
\small\bf e-mail: marek.jezabek@ifj.edu.pl
}
\vspace{0.5cm}
\date{\normalsize Dedicated to Stefan Pokorski on his 60th birthday\\
\small{\it To appear in Acta Physica Polonica B}}
\maketitle
\thispagestyle{empty}
\vspace{-4.5truein}
\begin{flushright}
{hep-ph/0205234}\\
{May 2002}
\end{flushright}
\vspace{4.0truein}
\begin{abstract}
A large hierarchy of the Dirac masses can result in a small hierarchy for 
the low energy masses of the active neutrinos. This can happen even if the 
Majorana masses of right-handed neutrinos are all equal. 
A realistic description 
of the observed neutrino masses and mixing can be obtained starting from 
a large hierarchy in the Dirac masses. A large mixing for solar neutrinos 
results from the neutrino sector. The small value of the MNS matrix element 
$U_{e3}$ is a natural consequence of the scheme. The masses of the two 
lighter neutrinos are related to the solar neutrino mixing angle: 
$\mu_1/\mu_2 = \tan^2\theta_\odot$.
\end{abstract}
%\PACS{PACS numbers come here}
\end{titlepage}

\section{Is there any mass hierarchy for active neutrinos?}

Let us start with the remark that the active neutrinos are exactly these
particles which experimentalists are studying. They couple to W and Z bosons.
There are three of them known as $\nu_e$, $\nu_\mu$ and $\nu_\tau$, and they
have very small masses which are reflected in mass scales governing neutrino
oscillations. In the oscillations of the solar and atmospheric neutrinos 
only differences\cite{dane,phenosol,phenoatm}
\begin{equation}
\Delta m^2_\odot = | \mu_2^2 - \mu_1^2 | \sim 5.0 \times 10^{-5} eV^2 
\end{equation}
\begin{equation}
\Delta m^2_{@}= | \mu_3^2 - \mu_2^2 | \sim 2.5 \times 10^{-3} eV^2 
\end{equation}
can be measured. The ratio of these two mass scales
\begin{equation}
\rho^{-1}_{exp} = \Delta m^2_{@}/ \Delta m^2_\odot \sim 50
\label{rhoexp}
\end{equation}
seems to provide a clear answer to the question asked in the title of this
section. Apparently yes. There is a hierarchy. However the correct answer
may be more subtle. Let us compare what Nature tells us through 
eq.(\ref{rhoexp}) with expectations based on a theory. The best theory of
neutrino masses we know is the see-saw mechanism \cite{seesaw}. It explains 
why the masses of the active neutrinos are much smaller than the masses of
all other fundamental fermions, i.e. charged leptons and quarks. The see-saw
mechanism implies that the masses of the active neutrinos are composite low
energy objects derived from more fundamental mass parameters. These more
fundamental masses are the Dirac masses describing couplings between
left-handed and right-handed neutrinos. and the Majorana masses of the
right-handed neutrinos. The right-handed neutrinos are singlets of the
standard model $SU_3\times SU_2\times U_1$ local gauge symmetry, so their
Majorana masses are not forbidden by gauge invariance. Majorana masses are 
not allowed for particles with non-zero electric charge. So, the masses of
charged leptons and quarks are all of the Dirac type and they all exhibit
a clear hierarchy
\begin{equation}
\matrix{
m_e \ll m_\mu \ll m_\tau  \cr
m_u \ll m_c \ll m_t  \cr
m_d \ll m_s \ll m_b  \cr}
\label{eq:4}
\end{equation}
If we assume that this hierarchical structure is a common feature of all
fundamental fermions, the Dirac masses of neutrinos should be also
hierarchical, i.e.
\begin{equation}
m_1 \ll m_2 \ll m_3  
\label{eq:5}
\end{equation} 
We still have to say something about the Majorana masses of the right-handed
neutrinos. The most natural thing is to assume that they are all equal.
So let us assume that there are three right-handed neutrinos and their
Majorana masses are equal to M:
\begin{equation}
M_R = M {\bf 1}
\label{eq:MR}
\end{equation}
Then the following sequence can be derived for the masses of three active
neutrinos:
\begin{equation}
\mu_1 = {m_1^2 \over M}\ ,  \qquad \mu_2 = {m_2^2 \over M}\ , \qquad 
\mu_3 = {m_3^2 \over M}\ .
\label{eq:7}
\end{equation}
If  $m_3/m_2 \sim m_t/m_c \sim 10^2$ is assumed, as suggested by many 
grand unified models, the ratio
\begin{equation} 
\rho^{-1}_{th} \approx {\mu_3^2 \over \mu_2^2} = 
\left( {m_3\over m_2}\right)^4 \sim 10^8
\label{rhoth}
\end{equation}
is obtained. When viewed from this perspective the hierarchy exhibited in 
eq.(\ref{rhoexp}) can be called a moderate one at best. It is much more
appropriate in fact to consider this small hierarchy as a small perturbation
of the situation without hierarchy.

\section{Reducing hierarchy}

Are we then forced to abandon the assumed hierarchy (\ref{eq:5}) of the
Dirac masses or the nice and economic postulate (\ref{eq:MR}) of equal
Majorana masses? Let us repeat the standard derivation of the mass formula
for the active neutrinos. Our guiding principle is to reduce the resulting
hierarchy as much as possible. The Dirac masses of neutrinos are described
by a $3\times 3$ matrix
\begin{equation}
N = U_R m^{(\nu)} U_L
\label{eq:9}
\end{equation}
with
\begin{equation}
m^{(\nu)} = {\rm diag}\left( m_1,m_2, m_3 \right)
\label{eq:10}
\end{equation}
As an unitary matrix $U_L$ cannot affect the resulting mass spectrum,
we assume
\begin{equation}
U_L = {\bf 1}
\label{eq:11}
\end{equation}
for simplicity. We may be led to reconsidering this when discussing the 
lepton mixing matrix.

The mass spectrum of the active neutrinos is given by a dimension five
operator $\cal N$. This operator is obtained as a low energy approximation
of a term resulting from the underlying renormalizable theory in the
next-to-leading order. The result is
\begin{equation}
{\cal N} = N^T M_R^{-1} N = {1\over M} {m^{(\nu)}}^T U_R^T U_R m^{(\nu)}
\label{eq:12}
\end{equation}
As the Majorana mass M in (\ref{eq:12}) is huge the resulting masses of 
the active neutrinos are small. The spectrum is extremely sensitive to 
the form of a symmetric unitary matrix
\begin{equation}
R = U_R^T U_R
\label{eq:13}
\end{equation}
so, the matrix $U_R$ plays a very important role in low energy physics and
its structure is imprinted in the masses of the active neutrinos.\footnote{
It is interesting to note that the analogous matrices for up and down
quarks play no role in low energy physics because they neither affect the
spectra of Dirac masses nor the electroweak charged currents.}
Unfortunately this mass spectrum is the only piece of information on
$U_R$ accessible at our low energies. So we have to guess some form of
$R$ and hope that the results obtained may to some extend justify our
cavalier attempt. $R = \bf 1$ is not acceptable because this would lead
us directly to the disastrous spectrum (\ref{eq:7}). Let us follow our
guiding principle and try to reduce the hierarchy of the resulting
spectrum as much as possible. Certainly
\begin{equation}
R = \left( \matrix{ 0  &   0  &   1  \cr
                    0  &   1  &   0  \cr
                    1  &   0  &   0  \cr } \right) = P_{13} 
\label{eq:14}
\end{equation}
seems to be a good candidate to achieve this goal. From 
(\ref{eq:12})-(\ref{eq:14})
\begin{equation}
{\cal N} = \mu
\left( \matrix{ 0  &   0  &   r  \cr
                    0  &   1  &   0  \cr
                    r  &   0  &   0  \cr } \right)
\label{eq:15}
\end{equation}
is obtained with $r = m_1 m_3/ m_2^2$ and $\mu = m_2^2 /M$. There is 
a doublet $(\nu_1, \nu_2)$ of mass $\mu r$ and a singlet $\mu_3$ of
mass $\mu$ in the spectrum resulting from (\ref{eq:15}). When this
spectrum is compared with those in (\ref{eq:7}) the reduction of 
hierarchy becomes evident. One may remark that this success is rather
problematic. If we want to interpret the mass splitting between singlet
and doublet as the origin of $\Delta m^2_{@}$ then $\Delta m^2_{\odot}$ 
is zero pushing our $\rho_{th}^{-1}$ to infinity, which seems to be
even worse than (\ref{rhoth}). We shall ignore this problem for a while.
It can be solved by introducing a small off-diagonal element in 
(\ref{eq:10}) removing mass degeneracy for $\nu_1$ and $\nu_2$ and
leading to non-zero $\Delta m^2_{\odot}$. These considerations dictate
ordering of eigenvalues after diagonalization of $\cal N$ which partly
fixes the form of a unitary matrix $O^\prime$ such that
\begin{equation}
{O^\prime}^T {\cal N} O^\prime = {\rm diag}
\left( \mu_1, \mu_2, \mu_3 \right) 
\label{eq:16}
\end{equation}
with $\mu_1 = \mu_2 = \mu r$, $\mu_3 =\mu$. The remaining freedom
will be removed completely by the perturbation splitting the masses of
$\nu_1$ and $\nu_2$. The result is
\begin{equation}
{O^\prime} = P_{23} U_{12}\left(\pm {\pi/ 4}\right)
\label{eq:17}
\end{equation}
with
\begin{equation}
P_{23} =
    \left( \matrix{ 1  &   0  &   0  \cr
                    0  &   0  &   1  \cr
                    0  &   1  &   0  \cr } \right)
\label{eq:18}
\end{equation}

\begin{equation}
U_{12}\left(\pm {\pi/ 4}\right) =
\left( \matrix{ 
{1\over\sqrt{2}}  &  \mp{i\over\sqrt{2}} &   0  \cr
{1\over\sqrt{2}}  &  \pm{i\over\sqrt{2}} &   0  \cr
                    0  &   0  &   1  \cr } \right)
\label{eq:19}
\end{equation}

\section{Lepton mixing matrix}

The Maki-Nakagawa-Sakata lepton mixing matrix\cite{MNS} can be expressed 
in terms of $O^\prime$ and $V_L$, where $V_L$ is a unitary matrix
diagonalizing $L^\dagger L$ and $L$ is the charged lepton mass matrix:
$ V_L L^\dagger L V_L^\dagger = {\rm diag}
\left( m_e^2, m_\mu^2, m_\tau^2 \right)$.
Then
\begin{equation}
\left( \matrix{ \nu_e    \cr
                \nu_\mu  \cr
                \nu_\tau \cr} \right) =
\left( \matrix{ 
 U_{e1} & U_{e2}  & U_{e3} \cr
 U_{\mu 1} & U_{\mu 2}  &  U_{\mu 3}   \cr
 U_{\tau 1}  & U_{\tau 2}  & U_{\tau 3}   \cr } \right)
\left( \matrix{ \nu_1    \cr
                \nu_2  \cr
                \nu_3 \cr} \right) =
U_{\rm MNS} \left( \matrix{ \nu_1    \cr
                \nu_2  \cr
                \nu_3 \cr} \right)
\label{eq:20}
\end{equation}
and
\begin{equation}
U_{\rm MNS} = V_L O^\prime = V_L P_{23} U_{12}\left(\pm {\pi/ 4}\right)
\label{eq:21}
\end{equation}
The structure in eq.(\ref{eq:21}) is striking. If $V_L$ is a matrix
with the element $(V)_{11} = 1$ and other non-zero elements in the
2-3 block the product $V_L P_{23}$ has the very same structure\footnote{
In general the product $V_L P_{23}$ is obtained from $V_L$ by exchanging
its second and third column. It may be considered as an efficient way to
ruin predictions associated with some special forms of $V_L$.
}. Moreover it is exactly this form of $V_L$ that can account for the
mixing of atmospheric neutrinos. Many authors considered lepton sector as 
the origin of maximal mixing for atmospheric neutrinos; see \cite{AFreview} 
and references therein. Particularly attractive models are based on
lopsided mass matrices \cite{ABB,AFreview}. So we do not spend more time 
on that problem because up to some irrelevant redefinitions\footnote{
Throughout this paper we ignore complex phases which are not important
for oscillations. Of course these phases are of crucial importance for 
$0\nu2\beta$ transitions.}
\begin{equation}
V_L P_{23} =
\left( \matrix{ 
1  & 0  &   0  \cr
0  &  {1\over\sqrt{2}}  &  {1\over\sqrt{2}}   \cr
0  & - {1\over\sqrt{2}}    &    {1\over\sqrt{2}} \cr } \right)
\label{eq:22}
\end{equation}
can be obtained following arguments of those papers. What we get from
(\ref{eq:21}) and (\ref{eq:22}) is known as the bi-maximal 
mixing\cite{bi-maxim}. It was a lot of fun to get this structure four
years ago. However, now the bi-maximal mixing is without any doubt
excluded by the experimental data \cite{phenosol}. Is this a problem
for the present scheme? Not really. The same perturbation which splits
the masses of $\nu_1$ and $\nu_2$ can push the solar mixing angle 
$\theta_\odot$ away from $\pm {\pi/4}$ in $U_{12}$. A perturbation
can be found producing $\Delta m^2_\odot$ and $\tan^2\theta_\odot$ in 
agreement with experiment \cite{JU02}.

\section{Can we test this picture ?}
The picture which we obtain is quite encouraging. Up to small
corrections the lepton mixing matrix can be written as
\begin{equation}
U_{\rm MNS} =
\left( \matrix{ 
1  & 0  &  0   \cr
0  & \cos\theta_{@}   & \sin\theta_{@}   \cr
0  & -\sin\theta_{@}   &  \cos\theta_{@}  \cr } \right)
\left( \matrix{ 
\cos \theta_\odot & \sin  \theta_\odot   &    0 \cr
-\sin\theta_\odot  & \cos\theta_\odot   &   0  \cr
0  &  0  &   1  \cr } \right)
\label{eq:23}
\end{equation}
This form explains the smallness of $U_{e3}$ in agreement with
CHOOZ limit \cite{CHOOZ} and SuperKamiokande data on atmospheric
$\nu_e$'s~\cite{dane,phenoatm}. Moreover, if the present picture
is correct there is a relation between $\mu_1$, $\mu_2$ and 
$\tan^2\theta_\odot$. Let us consider the 1-2 block of 
$P_{23}{\cal N}P_{23}$. For small off-diagonal elements in $m^{(\nu)}$,
c.f. eq(\ref{eq:10}) this $2\times 2$ matrix is proportional to
$$
\left( \matrix{ 
*  &  1 \cr
1  &  a \cr }\right) $$
with $|a|< 1$ and the element 1-1 small as a consequence of mass
hierarchy in $m^{(\nu)}$. Diagonalization of this sub-matrix produces
a unitary transformation in the 1-2 plane which is reflected 
in $U_{\rm MNS}$, see eq.({\ref{eq:23}). Thus the masses of $\nu_1$ and
$\nu_2$ are related to $\tan^2\theta_\odot$:
\begin{equation}
{\mu_1\over\mu_2} \approx \tan^2\theta_\odot
\label{eq:24}
\end{equation}
As a final remark let us note that the mass scale of the Majorana
masses is between $10^{10}$ and $10^{11}$ GeV if $m_2 \sim m_c$ 
is assumed. This range of Majorana masses may be quite interesting 
for baryogenesis; see \cite{jez01} and references therein.

\section{Acknowledgments}
I thank Piotr Urban for common work on the consequences of the scheme
presented here. I am very much indebted to Frans Klinkhamer
for a helpful discussion and suggesting the title of this article.

This work was done during my stay in the Institute f. Theoretische
Teilchenphysik, Universitaet Karlsruhe (TH).  
I would like to thank the Alexander-von-Humboldt Foundation for 
a grant which made my visit to Karlsruhe possible. 

This work is also supported in part by the KBN grant 5P03B09320 and 
the European Commission 5th Framework contract HPRN-CT-2000-00149.


\begin{thebibliography}{99}
\bibitem{dane}
T. Toshito [SuperKamiokande Collaboration], hep-ex/0105023; S. Fukuda et
al.[SuperKamiokande Collaboration], Phys. Rev. Lett. {\bf 86} (2001) 5651;
Phys. Rev. Lett. {\bf 86} (2001) 5656; Q.R. Ahmad et al. [SNO Collaboration]
Phys. Rev. Lett. {\bf 87} (2001) 071301; nucl-ex/0204009.
\bibitem{phenosol}
M.B. Smy, hep-ex/0202020; 
V. Barger, D. Marfatia, K. Whisnant, B.P. Wood, hep-ph/0204253; 
J.N. Bahcall, M.C. Gonzalez-Garcia, C. Pena-Garay,  hep-ph/0204314.
\bibitem{phenoatm}
G.L. Fogli, E. Lisi, A. Marrone, Phys.Rev. {\bf D65} (2002) 073028.  
\bibitem{seesaw}
M. Gell-Mann, P. Ramond and R. Slansky, in: P. van Nieuwenhuizen and
D.Z. Freedman (eds.), {\it Supergravity} (North Holland, Amsterdam, 1979)
p.315;\\
T. Yanagida , in: O. Sawada and A. Sugamote (eds.), {\it Proc. of the Workshop
on the Unified Theory and Baryon Number in the Universe} (KEK report 79-18,
1979) p.95;\\ 
S. Weinberg, Phys. Rev. Lett. {\bf 43} (1979) 1556;\\
R.N.~Mohapatra and G.~Senjanovic, {\it Phys. Rev. Lett.} {\bf 44}, 912 (1980).
\bibitem{MNS}
Z. Maki, M. Nakagawa and S. Sakata,
Prog.Theor.Phys. {\bf 23} (1960) 1174.
\bibitem{AFreview}
G. Altarelli and F. Feruglio, Phys.Rept. 320 (1999) 295; 
S.M. Barr and I. Dorsner, Nucl. Phys. {\bf B 585} (2000) 79;
I. Dorsner and S.M. Barr, Nucl. Phys. {\bf B 617} (2001) 493.
\bibitem{ABB}
K.S. Babu and S.M. Barr, Phys. Lett. {\bf B 381} (1996) 202;
C.H. Albright, K.S. Babu and S.M. Barr, Phys. Rev. Lett. {\bf 81} (1998) 116;
K.S. Babu, J.C. Pati and F. Wilczek, Nucl. Phys. {\bf B 566} (2000) 33.  
\bibitem{bi-maxim}
F. Vissani, hep-ph/9708483 (unpublished);
V. Barger, S. Pakvasa, T.J. Weiler and K. Whisnant, 
Phys. Lett. {\bf B437} (1998) 107;
A.J. Baltz, A.S. Goldhaber and M. Goldhaber, Phys. Rev. Lett. 
{\bf 81} (1998) 5730 ;
D.V. Ahluwalia 
Mod. Phys. Lett. {\bf A13} (1998) 2249; 
M. Je\.zabek and Y. Sumino, Phys. Lett. {\bf B 440} (1998) 327;
G. Altarelli and F. Feruglio,Phys. Lett. {\bf B 439} (1998) 112;
H. Georgi, S.L. Glashow, Phys.Rev. {\bf D 61} (2000) 097301  
(hep-ph/9808293).
\bibitem{JU02}
M. Je\.zabek and P. Urban, {\it A hidden hierarchy of neutrino masses},
to be published.
\bibitem{CHOOZ}
M. Apollonio et al. [CHOOZ Collaboration], Phys. Lett. {\bf B 466} 
(1999) 415.
\bibitem{jez01}
M. Je\.zabek, Nucl. Phys. (Proc.Suppl.) 100 (2001) 276.
\end{thebibliography}
\end{document}